\documentstyle[12pt]{article}
\makeatletter
\@addtoreset{equation}{section}
\makeatother

\begin{document}
\thispagestyle{empty}
\begin{flushright}              
BI-TP 99/22\\
IRB-TH-5/99\\
hep-ph/9910337\\
\end{flushright}
\vspace*{11mm}
\begin{center}
{\Large \bf Out of Equilibrium Thermal Field Theories}\\
{\Large \bf  - Finite Time after Switching on the Interaction}\\
{\Large \bf - Wigner Transforms of Projected Functions}\\
\vspace*{11mm}
{\large
 I.~Dadi\'c$^{1,2}$ }
\footnote{E-mail: dadic@faust.irb.hr}
\\[24pt]
$^1$ Ruder Bo\v{s}kovi\'{c} Institute,  Zagreb, Croatia\\
$^2$ Fakult\"at f\"ur Physik, Universit\"at Bielefeld,
Bielefeld, Germany \\
\vspace*{11mm}
\end{center}

\begin{abstract}
We study out of equilibrium thermal field theories with switching on
the interaction occurring at finite time using the Wigner transforms
of two-point functions.

For two-point functions we define the concept of projected function: 
it is zero if any of times refers to the time before switching on the 
interaction, otherwise it depends only on the relative coordinates. 
This definition includes bare propagators, 
one-loop self-energies, etc.
For the infinite-average-time limit of the Wigner
transforms of projected functions we define the analyticity assumptions:
(1) The function of  energy is analytic above (below) the real axis.
(2) The function goes to zero as the absolute value of energy approaches 
infinity in the upper (lower) semiplane.

Without use of the gradient expansion, we obtain the
convolution product of projected functions. 
We sum the Schwinger-Dyson series in closed form. In the calculation 
of the Keldysh component (both, resummed and single self-energy insertion 
approximation) contributions
appear which are not the Fourier transforms of projected  functions, 
signaling the limitations of the method.

In the Feynman diagrams there is no explicit
energy conservation at vertices, there is an overall
energy-smearing factor taking care of the uncertainty relations.

The relation between the theories with the
Keldysh time  path and with the finite time path enables one to rederive
the results, such as the cancellation of pinching, collinear, and 
infrared singularities, hard thermal loop resummation, etc.
\end{abstract}

\section{ Introduction}
Out of equilibrium thermal field theory
\cite{schwinger,keldysh}
has recently attracted considerable interest
\cite
{kadanoff,niemi,danielewicz,cshy,rammer,landsman,calzetta,eboli,remler,mleb,brown,gl,bio}.

In many applications one considers the
properties of the Green functions of almost equilibrated systems,
at infinite time after switching on the interaction. A recent approach
based on first principles has been successful in demonstrating
the cancellation of
collinear \cite{lebellac,niegawacom} and pinching singularities
\cite{as,altherr,bedaque,niegawa,gl2,id}, the extension of
the hard thermal loop (HTL) approximation \cite{pisarski,bp,frenkel}
to out of equilibrium
\cite{carrington,defu}, and applications to heavy-ion collisions
\cite{bdr,bdrs,bdrk}.
A weak point of the approach was that most of these
results were obtained under the assumption that the variation
of slow Wigner variables could be ignored or, in other words,
that these results were valid only in the lowest order of the
gradient expansion \cite{langreth,henning,bi}.

For some problems, e.g., heavy-ion collisions, both limitations are
undesired. One would like to consider large deviations from
equilibrium. One cannot wait infinitely long as these systems go 
apart after a very short
time, probably without reaching the stage of
equilibrium. In nuclear collisions,
short-time-scale features have been studied
in a number of papers
$\cite{grange,tohyama,reinhard,tohyama2,gwr,morawetz}$.

In the theories with switching on the interaction 
infinitely long before the
time of interest, one tries to get some information by 
extrapolation to early times.
However, in doing so, this information is either deformed or lost. Indeed,
relaxation phenomena include many processes that are expected
to terminate as one goes sufficiently far from the switching-on time of 
the interaction.
Thus one can expect that the full theory also describes early times.
In many papers, problems with finite switching-on
time$^{\cite{calzetta}}$,
especially those related to the inflatory phase of the
Universe (see Ref. \cite{boyanovsky} for further references) have been  
studied by
Wigner-transforming in relative distance and studying the dependence
on relative time directly$^{\cite{devega}}$. In such an approach,
one relies heavily on differential equations and numerical
methods.

In an attempt to remove weak points in both cases,
we suggest the application of the method of Wigner transform (but now 
also in relative time) to the case of switching on the interaction at 
finite time.The time-integration path C (see Fig. 1) is now closing 
the part of the real axis between $t_i$ (the switching-on time) and $t_f$ 
(the switching-off time); in the rest of the paper
the switching-off time is pushed to infinity.
If we push the switching-on time to minus infinity,
the connection to the Keldysh integration path is established.
It turns out that this connection is highly nontrivial.

To understand the limitations coming from the finite switching-on time,
we start with a generic two-point function $G(x,y)$.
The quantities $x$ and $y$
are four-vector variables with the time components in the range
$t_i<x_0,y_0<\infty$ (here $t_i$ is the time at which we switch on the
interaction; it is usually set to $-\infty $, but we set it to
$t_i=0$). We define the Wigner variables $s$ (relative 
space-time, relative variable) and
$X$ (average space-time, slow variable) as usual:
\begin{eqnarray}\label{wvi}
&&X={x+y \over 2},~s=x-y,
\cr
\nonumber\\
&&G(x,y)=G(X+{s \over 2},X-{s \over 2}).
\end{eqnarray}
The lower limit on $x_0$, $y_0$ implies the following conditions on $X_0$ and $s_0$:
$0<X_0,~-2X_0 < s_0 < 2X_0$. 
To define the projected function (truncated, ``mutilated function''
\cite{wiener}, PF in further text), we add two additional properties:
the function does not change with $\vec X$ (homogeneity assumption), 
it is a function
of ($s_0,\vec s$) within
the interval $-2X_0<s_0<2X_0$ and is identical to zero outside the interval:
\begin{eqnarray}\label{prfi}
&&F(X+{s\over 2}, X-{s\over 2})=\Theta(X_0)\Theta(2X_0-s_0)
\Theta(2X_0+s_0)\bar F(s_0,\vec s),
\cr\nonumber\\ &&
\bar F(s_0,\vec s)=\lim_{X_0\rightarrow\infty}F(X+{s\over 2},X-{s\over 2}).
\end{eqnarray}

Analogously, the Wigner transform of the projected function (WTPF) is
obtained from the Wigner transform of the function defined on the
infinite carrier of $s_0$ ($-\infty<s_0<\infty $) with the help
of the projection operators $P_{X_0}$, which are the Fourier
transforms of the above given $\Theta $'s:
\begin{eqnarray}\label{rftfp}
&&F_{X_0}(p_0,\vec p)=\int_{-\infty}^{\infty}dp'_0
P_{X_0}(p_0,p'_0)F_{\infty}(p'_0,\vec p),
\cr
\nonumber\\
&&P_{X_0}(p_0,p'_0)={1\over \pi}\Theta(X_0)
{\sin\left( 2X_0(p_0-p'_0)\right)\over p_0-p'_0},
~~\lim_{X_0\rightarrow\infty} P_{X_0}(p_0,p'_0)=\delta(p_0-p'_0),
\end{eqnarray}
where the subscript "$\infty$", as it appears in $F_{\infty}$ in (\ref{rftfp}) and
in other expressions of this paper, is the short
notation for the "$\lim_{X_0\rightarrow \infty}$".
As an illustration,
the Wigner transform of the convolution product of two-point functions
\begin{eqnarray}\label{pgfi}
C=A*B \Leftrightarrow C(x,y)=\int dz A(x,z)B(z,y)
\end{eqnarray}
is given by the gradient
expansion (note that we have
assumed the homogeneity in space coordinates,
which excludes any dependence on
$\vec X$):
\begin{eqnarray}\label{ge}
C_{X_0}(p_0,\vec p)=e^{-i\diamondsuit}A_{X_0}(p_0,\vec p)
B_{X_0}(p_0,\vec p),~~
\diamondsuit={1\over 2}(\partial_{X_0}^A\partial_{p_0}^B
-\partial_{p_0}^A\partial_{X_0}^B).
\end{eqnarray}
This general expression should be contrasted with
our result
valid for
$A$ and $B$ being projected functions (see Sec. 5):
\begin{eqnarray}\label{iffff}
C_{X_0}(p_0,\vec p)=\int dp_{01}dp_{02}
P_{X_0}(p_0,{p_{01}+p_{02}\over 2})
{1\over 2\pi}{ie^{-iX_0(p_{01}-p_{02}+i\epsilon)}
\over p_{01}-p_{02}+i\epsilon}
A_{\infty}(p_{01},\vec p)B_{\infty}(p_{02},\vec p).
\end{eqnarray}

The analytic properties of the
WTPF in the  $X_0\rightarrow \infty $ limit  as a
function of complex energy are very important for further analysis.
We define the following properties:
(1) the function of $p_0$ is analytic
above (below) the real axis, (2) the function  goes to zero
as $|p_0|$ approaches infinity in the upper (lower) semiplane.
The choice above (below) and upper (lower) refers to R (A)
components (note here that we do not require
such properties for the Keldysh components).
It is easy to recognize that the properties (1) and (2) are just the
definition of the retarded (advanced) function.
However, it is nontrivial, and not always true,
that the functions with the R (A) index satisfy them.

Under the assumption that $A$ or $B$ satisfies (1) and (2)
($A$ as advanced or $B$ as retarded)
Eq. (\ref{iffff}) can be integrated
even further. We obtain 
\begin{eqnarray}\label{ifff4}
C_{X_0}(p_0,\vec p)&=&\int dp'_{0}P_{X_0}(p_0,p'_{0})
A_{\infty}(p'_{0},\vec p)B_{\infty}(p'_{0},\vec p).
\end{eqnarray}
The convolution product of two two-point functions which are WTPF's
and satisfy (1) and (2) is also an WTPF. This product is then expressed
through the projection operator acting on a simple product of two WTPF's
given at $X_0^A=X_0^B=\infty $. 
In the $X_0\rightarrow\infty$ limit the convolution product of two WTPF's
which satisfy (1) and (2) is equal to the lowest-order contribution
in the gradient expansion. The result makes sense since moving
$X_0\rightarrow \infty$ is equivalent to $t_i\rightarrow -\infty$
at fixed $X_0$ in the standard approach.

We find that some quantities, obtained at low orders in
the perturbative expansion, e.g., bare propagators, one-loop
self-energies, belong to WTPF's.
This enables us to sum the
Schwinger-Dyson series with the propagators and self-energies 
as both these quantities are
WTPF. Under the conditions (1) and (2), the retarded, advanced, and
Keldysh components at finite $X_0$ are obtained by a simple
action (smearing)
of the projection operator onto the corresponding quantities obtained
at $X_0=\infty $, and the convolution product is a simple
multiplication:
\begin{eqnarray}\label{ifff5}
&&{\cal G}_{R,X_0}(p_0,\vec p)=\int dp_{01}P_{X_0}(p_0,p_{01})
{\cal G}_{R\infty}(p_{01},\vec p),\cr
\nonumber\\
&&{\cal G}_{R,\infty}(p_{01},\vec p)={G_{R,\infty}(p_{01},\vec p)\over
1-i\Sigma_{R,\infty}(p_{01},\vec p)G_{R,\infty}(p_{01},\vec p)},
\end{eqnarray}
and similarly for the advanced component. The calculation of the
Keldysh component requires a more elaborate treatment: one reduces the multiplication
to a double (a single in some cases) convolution product; the result
contains terms that are non-WTPF terms.
Also, the one-loop contribution to the Keldysh component 
is corrected by non-WTPF terms.

Now, it is ${\cal G}_{R,\infty}(p_0,\vec p)$ as given in Eq.
(\ref{ifff5})
(and similar expressions for the advanced and Keldysh components and the
single self-energy insertion approximation to ${\cal G}_{R(A,K),\infty}$)
to which previous results on the cancellation of  pinching
singularities$^{\cite{as,altherr,niegawa,id}}$
and the HTL resummation$^{\cite{carrington,defu}}$
(and also on the cancellation of collinear$^{\cite{lebellac,niegawacom}}$
and infrared singularities if the properties (1) and (2)
hold at the two-loop level) apply.

From our study one can deduce a general rearrangement of
the perturbation expansion at the non-Keldysh integration path:
the contributions look like the zeroth order of the gradient expansion
with the slow coordinate ($X_0$) pushed to $+\infty $, but
the use of the PF manifests itself as the
appearance of the $(\sum_j q_{0j}+i\epsilon)^{-1}$
factor instead of $-\pi\delta(\sum_j q_{0j})$ at each vertex,
and as an overall projection (smearing) operator instead of the exact
conservation of energy.

Our study suggests that the results obtained by using the
Keldysh integration path ($t_i\rightarrow -\infty $) could be
related to the results of our
approach ($t_i$ finite). This relation is possible at low orders
of the perturbation expansion, i.e., as long as the expressions
involved are the projected functions not breaking assumptions (1)
and (2). Technically, the amplitudes are related by (\ref{rftfp}),
where  
"$F_{X_0}$" is the contribution in our approach ($t_i$ finite)
and "$F_{\infty} $" is substituted by the corresponding lowest-order contribution in
the gradient expansion in the theories with $t_i\rightarrow -\infty $.

Finally, we note that our method is helpful in problems related to the time
evolution of the system. Additional problems related to the gradient
expansion in space components, appearing together with 
space inhomogeneity, will not benefit from our method.

The paper is organized as follows.
In Sec. 2 we give a general setup of out of equilibrium
thermal field theory.
In Sec. 3 we define finite-time Wigner transforms,
define the projection operators, and
introduce the notion of projected functions. We define analyticity
assumptions (1) and (2).
In Sec. 4 we define a few important examples of projected functions namely
bare propagators and one-loop self-energies,  and find that they satisfy the
analyticity
assumptions (1) and (2).
In Sec. 5 we analyze the properties of
the product of two and $n$ two-point functions.
We discuss the reduction of the inverse bare propagator to the
space of projected functions.
In Sec. 6 these properties are used to study the product of two pole
contributions, and to discuss the reduction of the inverse bare propagator
to the space of projected functions and the equations of motion.
We sum the Schwinger-Dyson series
for retarded, advanced, and Keldysh components of the propagator.
We discuss the appearance of pinching singularities in our scheme.
Section 7 is devoted to the modifications of Feynman rules in 
coordinate and momentum space. It is indicated that, in the absence of
breakdown of assumptions (1) and (2), all energy denominators
appearing at vertices can be replaced by delta functions.
Section 8 is a summary of the results and ideas described in the
paper.
\section{Out of Equilibrium Setup}
We start by assuming that the system has been prepared at
some initial time $t_i=0$ (To avoid inessential complications, we
assume that the zero-temperature renormalization has
already been performed).
At $t_i$ the interaction is switched on and at time $t_f$
it is switched off (we shall take the limit $t_f\rightarrow \infty $).
For $t_i<t_1,t_2<t_f$, the system evolves under the evolution
operator \cite{landsman}
\begin{eqnarray}\label{u12}
U(t_2,t_1)=T_c[\exp i\int_c d^4x'{\cal L}_I(x')],
\end{eqnarray}
where $c$ is the integration contour connecting $t_1$ and $t_2$
in the complex time plane and $T_c$ is the contour ordering operator.
We provide $T_c$ with an extra property:
for all times not belonging to the contour it gives zero.

The Heisenberg field operator $\phi(x)$ is obtained from the free field
$\phi_{in}(x)$ in the interaction picture as
\begin{eqnarray}\label{ffield}
\phi(x)=U(t_i,t)\phi_{in}(x)U(t,t_i),
\end{eqnarray}
%
\begin{eqnarray}\label{tffield}
\phi(x)=T_C[\phi_{in}(x)\exp i\int_C d^4x'{\cal L}_I(x')],
\end{eqnarray}
where all fields on the right-hand side are in the interaction picture,
and C is the integration contour of Fig. 1 (with
the switching-off time pushed to infinity, $t_f\rightarrow +\infty $).
In the Heisenberg picture, the average values of the operators
are obtained as
\begin{eqnarray}\label{av}
<O(t)>=Tr \rho O(t),
\end{eqnarray}
where $\rho $ is the density operator admitting the Wick decomposition.
Specially, we define the two-point Green function as
\begin{eqnarray}\label{ig}
G^{(C)}(x,x')=-i<T_C\phi(x)\phi(x')>.
\end{eqnarray}
With the help of (\ref{u12}) it can be written as
\begin{eqnarray}\label{iig}
G^{(C)}(x,x')=-i<T_C\left(\exp
[i\int_C d^4x'{\cal L}_I(x')]\phi_{in}(x)\phi_{in}(x')\right)>.
\end{eqnarray}

In Eq. (\ref{iig}) it is implicitly assumed that the 
interaction Lagrangian
does not depend on time explicitly. 
Indeed, direct time dependence through
time-dependent perturbation, or through the background field which
depends directly on time, $\psi(x,t)=\phi(t)+\bar \psi(x,t)$, would break
our scheme through the appearance
of two-point functions which are not
projected functions.

We assume the single-particle density operator to be stationary
with respect to the free Hamiltonian
 $H_0=\sum_jH_j$ (for an alternative choice of
the initial density, see Ref. \cite{baacke}):
\begin{eqnarray}\label{do}
\rho=\sum_n|\psi_n>p_n<\psi_n|={1\over Z}\exp (-\sum_{j}\beta_jH_j),
\end{eqnarray}
where the "temperature" function$^{\cite{calzetta}}$ $\beta_j$
(the ``temperature'' of the $j^{th}$ degree of freedom)
is adopted to obtain the given initial-state particle distribution.
Now, one has
\begin{eqnarray}\label{bh}
\sum_j\beta_jH_j=
\int d^4p\beta(p_0)p_0\Theta(p_0)\delta(p_0^2-\vec p^2-m^2)a^+_pa_p,
\end{eqnarray}
and obtains the distribution function $f(p_0)$ as a
function of $\beta(p_0)$,
\begin{eqnarray}\label{fpo}
f_{\beta}(p_0)={1\over p_0}Tr p_0\rho ={1 \over \exp \beta(p_0)p_0\mp1},
\end{eqnarray}
or the inverse relation
\begin{eqnarray}\label{fbeta}
\beta_f(p_0)={\log ({1\over f(p_0)}\pm 1) \over p_0}.
\end{eqnarray}
The free fields are expanded in creation and annihilation operators as
\begin{eqnarray}\label{field}
\phi(x)=\int {d^3p\over (2\pi)^32\omega_p}(a_pe^{-ipx}+a^+_pe^{ipx}),
\end{eqnarray}
with $p_0=\omega_p=(\vec p^2+m^2)^{1/2}$:
\begin{eqnarray}\label{oper}
&&<a^+_pa_{p'}>=(2\pi)^32\omega_pf(\omega_p)\delta (\vec p-\vec p'),
\cr
\nonumber\\
&&
<a_pa^+_{p'}>=(2\pi)^32\omega_p(1\pm f(\omega_p))
\delta (\vec p-\vec p'),
\end{eqnarray}
where $f(\omega_p)$ is the given initial distribution. 

For completeness, in Eqs. (\ref{bh})-(\ref{oper}) we include free fermions
(lower case) in addition to free 
bosons (upper case).  
In the case of
spin-1/2, spin-1,  or higher-spin particles, additional spinor or 
tensor indices must appear.
For simplicity, we do not show them explicitly. 

The noninteracting contour Green function is given as 
\begin{eqnarray}\label{nonig}
G^{(C)}_{in}(x,x')=-i<T_C\phi_{in}(x)\phi_{in}(x')>.
\end{eqnarray}
Depending on whether the times $x_0$ and $x'_0$ belong to the
upper (``1'') or lower (``2'') part of the path $C$, the function
$G^{(C)}_{in}(x,x')$ splits into the components
$G_{\mu,\nu,in}(x,x'),~~~~~\mu,\nu=1,2$. For the times
$x_0<0$ or $x'_0<0$,
the Green function is equal to zero owing to our definition of $T_C$.

\section{ Projected Functions}
Let us start with the two-point function $G(x,y)$.
The quantities $x$ and $y$
are four-vector variables with time components in the range
$t_i<x_0,~y_0<\infty$ (here $t_i$ is the time at which we switch on the
interaction; it is usually set to $-\infty $, but we set it to
$t_i=0$). We define the Wigner variables as usual:
\begin{eqnarray}\label{wv}
&&X={x+y \over 2},~s=x-y,
\cr
\nonumber\\
&&G(x,y)=G(X+{s \over 2},X-{s \over 2}).
\end{eqnarray}
The lower limit on $x_0$, $y_0$ implies
$0<X_0,~-2X_0 < s_0 < 2X_0$. The values of the function $G$ for the $(X,s)$ not
satisfying these conditions are physically irrelevant. Our definiton of
time ordering operator (see Sec II) sets them to zero, so that we can rewrite
Eq. (\ref{wv}) as
\begin{eqnarray}\label{wvt}
G(x,y)=\Theta(X_0)\Theta(2X_0-s_0)\Theta(2X_0+s_0)\bar G(X+{s \over 2},X-{s \over 2}).
\end{eqnarray}
Note here that 
the function $\bar G$ defined by Eq. (\ref{wvt}) in general depends on $X_0$. 
At the points $(X,s)$ which do not belong to the carrier of projection operator, the values of
$\bar G$ are arbitrary. This freedom is used to define projected functions.
The two-point function can be expressed in 
terms of the Wigner transform (i.e. Fourier transform with respect to $s_0, s_i$):
\begin{eqnarray}\label{ft}
G(X+{s \over 2},X-{s \over 2})=(2\pi)^{-4}
\int d^4pe^{-i(p_0s_0-\vec p\vec s)}G(p_0,\vec p;X).
\end{eqnarray}
Here
\begin{eqnarray}\label{ift}
&&G(p_0,\vec p;X)=\int_{-2X_0}^{2X_0}ds_0\int d^3s
e^{i(p_0s_0-\vec p\vec s)}G(X+{s \over 2},X-{s \over 2})
\cr\nonumber\\&&
=\int_{-\infty}^{\infty}\int ds_0\int d^3s
e^{i(p_0s_0-\vec p\vec s)}\Theta(X_0)\Theta(2X_0-s_0)\Theta(2X_0+s_0)\bar G(X+{s\over 2},X-{s\over 2}).
\end{eqnarray}
We adopt a simplifying assumption of the homogeneity in space 
coordinates. This 
assumption excludes any dependence on $\vec X$ and we drop 
it as an argument of the function.

The product of $\Theta$ functions is a projection operator with a simple Fourier
transform 
\begin{eqnarray}\label{sftf}
P_{X_0}(p_0,p'_0)={1\over 2\pi}\Theta(X_0)\int_{-2X_0}^{2X_0}ds_0e^{is_0(p_0-p'_0)}
={1\over \pi}\Theta(X_0){\sin\left( 2X_0(p_0-p'_0)\right)\over p_0-p'_0},
\end{eqnarray}
and
\begin{eqnarray}\label{isftf}
e^{-is_0p'_0}\Theta(X_0)\Theta(2X_{0}+s_{0})\Theta(2X_{0}-s_{0})
=\int dp_0e^{-is_0p_0}P_{X_0}(p_0,p'_0).
\end{eqnarray}
It is important to note that
\begin{eqnarray}\label{dsftf}
\lim_{X_0\rightarrow \infty}P_{X_0}(p_0,p'_0)
=\lim_{X_0\rightarrow \infty}{1\over \pi}
{\sin\left( 2X_0(p_0-p'_0)\right)\over p_0-p'_0}
=\delta(p_0-p'_0).
\end{eqnarray}
There is a
hierarchy of the $P_{X_0}$ projectors:
\begin{eqnarray}\label{sss}
P_{X_{0,M}}(p_0,p"_0)=\int dp'_0P_{X_0}(p_0,p'_0)P_{X'_0}(p'_0,p"_0),
~~X_{0,M}=min(X_0,X'_0).
\end{eqnarray}

In this paper, the projected function is a very special
two-point function $F(x,y)=F(X+s/2,X-s/2)$: it does not depend on $\vec X$, 
it is a function
of ($s_0,\vec s$) within
the interval $-2X_0<s_0<2X_0$ and identical to zero outside:
\begin{eqnarray}\label{prf}
F(X+{s\over 2}, X-{s\over 2})=\Theta(X_0)\Theta(2X_0-s_0)\Theta(2X_0+s_0)\bar F(s_0,\vec s).
\end{eqnarray}
Function $\bar F$ is related to the limit $X_0\rightarrow\infty $:
\begin{eqnarray}\label{prfl}
\lim_{X_0\rightarrow\infty}F(X+{s\over 2}, X-{s\over 2})=\bar F(s_0,\vec s).
\end{eqnarray}
Important property of the projected function 
is that the whole $X_0$ dependence is introduced by the projection operator.
\begin{eqnarray}\label{rftf}
F_{X_0} (p_0,\vec p)=[P_{X_0}F_{\infty}](p_0,\vec p)=\int_{-\infty}^{\infty}dp'_0
P_{X_0}(p_0,p'_0)F_{\infty}(p'_0,\vec p),
\end{eqnarray}
Important examples of projected functions are retarded, advanced,
and Keldysh components of free propagators. Further examples
will emerge in the next sections.

For further analysis, the analytic properties of the 
$X_0\rightarrow \infty$ limit of the WTPF as a function of complex energy
are very important. We define the following properties: (1) the
function of $p_0$ is analytic above (below) the real axis, (2) the 
function goes to zero as $|p_o|$ approaches infinity in the upper 
(lower) semiplane. The choice above (below) and upper (lower)
refers to R(A) components.

One should note that the properties of the projected functions are tightly
related to the abrupt cutoff at $t_i$. Any smoothing of the cutoff
would also change these properties.

\section{Examples of Projected Functions}
\subsection{Poles in the Energy Plane}
We start with the simplest projected functions: simple poles
in the energy plane.
The pole contribution to the Green function is
\begin{eqnarray}\label{p0}
{\cal G}_{\infty,pole}(p_0)={a\over p_0-\bar p_0}.
\end{eqnarray}
For $Im\bar p_0<0$, it
satisfies assumptions (1) and (2)
as a retarded component, but not as an advanced component
(and for $Im\bar p_0>0$, just the opposite).

It can be projected to finite $X_0$:
\begin{eqnarray}\label{p1}
{\cal G}_{X_0,pole}(p_0)=a{1-e^{-2iX_0(p_0-\bar p_0)
sgn(Im\bar p_0)}\over p_0-\bar p_0}.
\end{eqnarray}
For any finite $X_0$, the function ${\cal G}_{X_0,pole}(p_0)$ is regular at
$\bar p_0$.
For large $X_0 $, Eq. (\ref{p1}) exhibits the exponential decay
$e^{-2X_0|Im\bar p_0|}$ independently of the sign of $Im\bar p_0$.

It can be transformed back to the variables $(X_0, s_0)$:
\begin{eqnarray}\label{p2}
{\cal G}_{pole}(X_0+{s_0\over 2},X_0-{s_0\over 2})=iae^{-i\bar p_0s_0}
\left(\Theta(s_0 sgn(Im\bar p_0))
-\Theta(-s_0 sgn(Im\bar p_0)-2X_0)\right).
\end{eqnarray}
Evidently, this contribution is a projected function. For $Im\bar p_0<0$,
it is different from zero only at $0<s_0<2X_0$, i.e.,
it is a retarded function, and for $Im\bar p_0>0$, it is 
an advanced function.

\subsection{ Propagator}
We start with Eqs. (\ref{field}), (\ref{oper}), and (\ref{nonig}).
The transition to the R/A basis is straightforward.
Careful calculation gives for the retarded component ($0<x_0$, $0<y_0$):
\begin{eqnarray}\label{nonigrr}
G_R(x,y)=-G_{1,1}+G_{2,1}=\int d^4p
{-i\over p^2-m^2+2i\epsilon p_0}e^{-ip(x-y)},
\end{eqnarray}
and for the Keldysh component:
\begin{eqnarray}\label{nonigrk}
G_K(x,y)=G_{1,1}+G_{2,2}=
\int d^4p
2\pi\delta(p^2-m^2)(1\pm 2f(\omega_p))e^{-ip(x-y)}.
\end{eqnarray}
As our $G_R$ and $G_K$ depend only on $s=x-y$ and
vanish at times before switching on
 the interaction, they are projected functions.
The Wigner transform over the infinite $x_0-y_0$ interval gives as
usual$^{\cite{niemi,mleb,bio}}$ (note, however, that we
avoid$^{\cite{id}}$ using the nonanalytic function $\epsilon(p_0)$
in the expression for $G_{K,\infty}$):
\begin{eqnarray}\label{fnonigrkr}
&&G_{R,\infty}(p)={-i\over p^2-m^2+2i\epsilon p_0},
\cr
\nonumber\\
&&G_{K,\infty}(p)=-(1\pm 2f(\omega_p))
\omega_p^{-1}\left(p_0G_{R,\infty}(p)-p_0G_{A,\infty}(p)\right).
\end{eqnarray}
The $\epsilon$ parameter, which regulates these expressions, 
should be kept uniformly finite
during the calculations, and the limit $\epsilon \rightarrow 0$ 
should be taken last of 
all$^{\cite{landsman}}$. This specially means that 
$\lim_{X_0\rightarrow\infty}\exp(-X_0\epsilon)=0$ 
and the terms containing this factor vanish in the 
$X_0\rightarrow\infty$ limit.

The finite Wigner transform ($x_0>0, y_0>0,X_0=(x_0+y_0)/2$)
is obtained by smearing
\begin{eqnarray}\label{ffkr}
G_{R,X_0}(p)=[P_{X_0}G_{R,\infty}](p)=-G^
*_{A,X_0}(p),~G_{K,X_0}(p)=
[P_{X_0}G_{K,\infty}](p).
\end{eqnarray}
It is easy to verify that neither the spinor nor the tensor factor changes
our conclusion (\ref{ffkr}).
One can even integrate expression (\ref{ffkr}). For a scalar
particle, one obtains
\begin{eqnarray}\label{ffkrs}
&&G_{R,X_0}^0(p)={-i\over p_0^2-\vec p^2-m^2+2i\epsilon p_0}
\left(1-(\cos 2X_0\omega_p -i{p_0\over\omega_p}\sin 2X_0\omega_p)
e^{2iX_0(p_0+i\epsilon)}\right)
\cr
\nonumber\\
&&=G_{R,\infty}^0(p_0)
\left(1-(\cos 2X_0\omega_p -i{p_0\over\omega_p}\sin 2X_0\omega_p)
e^{2iX_0(p_0+i\epsilon)}\right).
\end{eqnarray}
It is important to observe that, at any finite $X_0$,
the above expression is
not singular at $p_0=\pm\omega_p$.

Evidently, for $X_0 \rightarrow \infty $, the first term in
$G_{R,X_0}(p)$ gives $G_{R,\infty }$, while the other two
"oscillate out".
For the Keldysh component, one needs
\begin{eqnarray}\label{pgrx}
&&(p_0G_{R}^0)_{X_0}(p)=p_0G_{R,\infty}^0(p_0)
\left(1-(\cos 2X_0\omega_p -i{\omega_p\over p_0}\sin 2X_0\omega_p)
e^{2iX_0(p_0+i\epsilon)}\right).
\end{eqnarray}
Then one can use the analogy to (\ref{fnonigrkr}):
\begin{eqnarray}\label{ffkkx}
G_{K,X_0}^0(p)=-(1\pm 2f(\omega_p))
\omega_p^{-1}\left((p_0G_{R}^0)_{X_0}(p)-(p_0G_{A}^0)_{X_0}(p)\right).
\end{eqnarray}
For a spinor particle, one obtains
\begin{eqnarray}\label{ffkrsp}
G_{R,X_0}^{1/2}(p)=
i\gamma^0e^{2iX_0(p_0+i\epsilon)}
(\omega_p-{p_0^2\over\omega_p})\sin 2X_0\omega_p +
G_{R,X_0}^{o}(p)(\gamma^{\mu}p_{\mu}+m).
\end{eqnarray}
Similarly, for a vector particle (for simplicity, we choose the
Feynman gauge):
\begin{eqnarray}\label{ffkrv}
&&G^1_{\mu,\nu,R,X_0}(p)=g_{\mu,\nu}G_{R,X_0}^{0}(p).
\end{eqnarray}
We note here that the explicit expressions (\ref{ffkrs})-(\ref{ffkrv})
will not be necessary for further discussion.
\subsection{ One-Loop Self-Energy}
To discuss the amputated one-loop self-energy, we start with 
(the underlying theory includes bosons and fermions with three
point vertices,
but spin and internal
symmetry indices are suppressed for simplicity of presentation)
\begin{eqnarray}\label{se}
&&\Sigma(x,y)=g^2S(x,y)D(x,y)
\cr
\nonumber\\
&&=g^2\int d^4pe^{-ips}P_{X_0}(p_0,p'_0)S_{\infty}(p'_0,\vec p)
\int d^4qe^{-iqs}P_{X_0}(q_0,q'_0)D_{\infty}(q'_0,\vec q).
\end{eqnarray}
The Wigner transform (with respect to $s=x-y$) is
\begin{eqnarray}\label{ise}
&&\Sigma_{X_0}(p_{01},\vec p_1)
=g^2\int_{-2X_0}^{2X_0} ds_0\int dp_0d^3pdq_0
e^{-i(p_0+q_0)s_0}\Theta(2X_0-s_0)\Theta(2X_0+s_0)
\cr
\nonumber\\
&&\times
S_{\infty}(p'_0,\vec p)D_{\infty}(q'_0,\vec p_1-\vec p)
\cr
\nonumber\\
&&=g^2\int dp_0'd^3pdq_0'P_{X_0}(p_{01},p_0'+q_0')
S_{\infty}(p'_0,\vec p)D_{\infty}(q'_0,\vec p_1-\vec p),
\end{eqnarray}
where as an intermediary step we have used the representation of the
bare propagators (\ref{ffkr}) and the representation of the
projectors (\ref{sftf}) and (\ref{isftf}). Finally, one reads
(\ref{ise}) in the R/A basis 
($\Sigma_{R(A)}=-(\Sigma_{11}+\Sigma_{21(12)})$, 
$\Sigma_K=-(\Sigma_{11}+\Sigma_{22})$), as
\begin{eqnarray}\label{fskr}
\Sigma_{R(A),X_0}(p)=[P_{X_0}\Sigma_{R(A),\infty}](p),~
\Sigma_{K,X_0}(p)=[P_{X_0}\Sigma_{K,\infty}](p).
\end{eqnarray}
To calculate $\Sigma_{R,\infty}$ and $\Sigma_{K,\infty}$ we start
with Eqs. (II.23-25) in \cite{id}.
After taking into account that the product of the retarded with the advanced
function, with the same time variables, vanishes, one obtains
\begin{eqnarray}\label{sigmar}
\Sigma_{R,\infty }(q)={ig^2\over 2}\int {d^4k\over (2\pi)^4}
(h(k_0,\omega_k)+h(q_0-k_0,\omega_{q-k}))D_{R,\infty}(k)S_{R,\infty}(q-k)F,
\end{eqnarray}
where $D$ and $S$ are bare scalar propagators,
$h(k_0,\omega_k)=-k_0\omega_k^{-1}(1\pm 2f(\omega_p))$, and
the factor $F=F(k_0,|\vec k|,q_0,|\vec q|,\vec k\vec q,...)$
includes the information about spin and internal degrees of freedom
($F=1$ if all particles are scalars).
Now, to verify that the one-loop self-energy
$\Sigma_{R,\infty} $ satisfies assumptions (1) and (2), one observes
that the vacuum contribution satisfies them (for exceptions, see, e.g.,
Refs. \cite{henning,henning2}), while the
contributions to $\Sigma_{R,\infty} $ from various $k_0$ points are
linear and additive in
distribution functions.

For finite $\epsilon $, this contribution possesses singularities
only below the real axis in the complex $q_0$ plane, and vanishes as
$|q_0|\rightarrow\infty$
in the upper semiplane.
However, there is no guarantee that
the imaginary part of $\Sigma_{R,\infty}$ is negative.

Using the same method one calculates the Keldysh component.
Although assumptions (1) and (2) are not imposed on $\Sigma_K $,
one can decompose this component into two pieces 
$\Sigma_K=-\Sigma_{K,R}+\Sigma_{K.A}$: 
\begin{eqnarray}\label{omega0r}
\Sigma_{K,R(A),\infty}(q)=\mp{ig^2\over 2}\int {d^4k\over (2\pi)^4}
(1+h(k_0,\omega_k)h(q_0-k_0,\omega_{q-k}))
D_{R(A),\infty}(k)S_{R(A),\infty}(q-k)F,
\end{eqnarray}
where $\Sigma_{K,R(A)}$ satisfies assumptions (1) and (2) in the way the retarded
(advanced) function does. General analytic properties of the expressions
of the type (\ref{sigmar}) and (\ref{omega0r}) are well known: there
is a discontinuity (cut) along the real axis, starting at thresholds
for real processes and extending to $\pm \infty $.
\section{Convolution Product of Two Two-Point Functions }
Let us now consider the convolution product of two Green functions:
\begin{eqnarray}\label{pgf}
C=A*B \Leftrightarrow C(x,y)=\int dz A(x,z)B(z,y).
\end{eqnarray}
In terms of Wigner transforms:
\begin{eqnarray}\label{pgff}
&&C(p_0,\vec p;X)=\int_{-2X_0}^{2X_0}ds_0\int d^3s\int d^4z
e^{i(p_0s_0-\vec p~\vec s)}\cr
\nonumber\\
&&
\times{1\over (2\pi)^4}\int d^4p_1
e^{-i(p_{01}s_{01}-\vec p_1 \vec s_1)}A(p_{01},\vec p_1;X_1)\cr
\nonumber\\
&&
\times{1\over (2\pi)^4}\int d^4p_2
e^{-i(p_{02}s_{02}-\vec p_2 \vec s_2)}B(p_{02},\vec p_2;X_2),
\cr
\nonumber\\
&&X_1=X+{s_2\over 2},~X_2=X-{s_1\over 2},~s_1=x-z,~s_2=z-y.
\end{eqnarray}
The assumed translational invariance helps us easily integrate
the space components of momenta and coordinates. To do so, we substitute 
$d^3\vec sd^3\vec z$ by $d^3\vec s_1d^3\vec s_2$ (Jacobian $J=1$)
\begin{eqnarray}\label{sz}
\vec s=\vec s_1+\vec s_2,~\vec z={-\vec s_1+\vec s_2\over 2}+\vec X.
\end{eqnarray}
The momenta should be equal ($\vec p=\vec p_1=\vec p_2$) and one obtains
(note that the dependence on $X$'s is reduced to the dependence
on $X_0$'s; further in the text it is indicated as an index),
\begin{eqnarray}\label{pgsz}
&&C_{X_0}(p_0,\vec p)={1\over (2\pi)^2}\int_{-2X_0}^{2X_0}ds_0\int
dz_0\int dp_{01}\int dp_{02}\cr
\nonumber\\
&&
\times e^{i(p_0s_0-p_{01}s_{01}-p_{02}s_{02})}A_{X_{01}}(p_{01},\vec p)
B_{X_{02}}(p_{02},\vec p).
\end{eqnarray}
For energy integrals, we proceed in a somewhat different way. We
shrink our choice of the functions $A(x,z)$ and $B(z,y)$ to the
projected functions. Then we can use the connection to the
Wigner transforms on the infinite carrier:
\begin{eqnarray}\label{pgszff}
&&C_{X_0}(p_0,\vec p)={1\over (2\pi)^2}\int_{-2X_0}^{2X_0}ds_0
\int dz_0\int dp_{01}\int dp_{02}
e^{i(p_0s_0-p_{01}s_{01}-p_{02}s_{02})}\cr
\nonumber\\
&&
\times\int dp'_{01} P_{X_{01}}(p'_{01},p_{01})A_{\infty}(p_{01},\vec p)
\int dp'_{02} P_{X_{02}}(p'_{02},p_{02})B_{\infty}(p_{02},\vec p).
\end{eqnarray}
The integration $dp'_{01}dp'_{02}$ is easily performed with the help of
Eq. (\ref{isftf})and one obtains
\begin{eqnarray}\label{pgszfff}
&&C_{X_0}(p_0,\vec p)={1\over (2\pi)^2}\int_{-2X_0}^{2X_0}ds_0
\int dz_0\int dp_{01}\int dp_{02}
e^{i(p_0s_0-p_{01}s_{01}-p_{02}s_{02})}\cr
\nonumber\\
&&
\times\Theta(2X_{01}+s_{01})\Theta(2X_{01}-s_{01})
\Theta(2X_{02}+s_{02})\Theta(2X_{02}-s_{02})
A_{\infty}(p_{01},\vec p)B_{\infty}(p_{02},\vec p).
\end{eqnarray}
The product of $\Theta $ functions is transformed into
$\Theta(2X_{0}+s_{0})\Theta(2X_{0}-s_{0})\Theta(z_0)$. Then
\begin{eqnarray}\label{fff}
C_{X_0}(p_0,\vec p)=\int \int dp_{01}dp_{02}\delta(p_{0},p_{01},p_{02})
A_{\infty}(p_{01},\vec p)B_{\infty}(p_{02},\vec p).
\end{eqnarray}
Here
\begin{eqnarray}\label{ddd}
&&\delta(p_{0},p_{01},p_{02})={1\over (2\pi)^2}\int_{-2X_0}^{2X_0}ds_0
\int_0^{\infty}dz_0
e^{i(p_0s_0-p_{01}s_{01}-p_{02}s_{02})}\cr
\nonumber\\
&&={1\over (2\pi)^2}\int_{-2X_0}^{2X_0}ds_0
\int_0^{\infty}dz_0
e^{i(s_0(p_0-{p_{01}+p_{02}\over2})+(z_0-X_0)(p_{01}-p_{02}+i\epsilon))}
\cr
\nonumber\\
&&=P_{X_0}(p_0,{p_{01}+p_{02}\over2})
{1\over 2\pi}{i \over p_{01}-p_{02}+i\epsilon}
e^{-iX_0(p_{01}-p_{02}+i\epsilon)},
\end{eqnarray}
where we have used
\begin{eqnarray}\label{zint}
\int_0^{\infty}dz_0e^{iz\alpha}={i\over \alpha+i\epsilon}.
\end{eqnarray}
We can write the final expression as
\begin{eqnarray}\label{ffff}
C_{X_0}(p_0,\vec p)=\int dp_{01}dp_{02}
P_{X_0}(p_0,{p_{01}+p_{02}\over 2})
{1\over 2\pi}{ie^{-iX_0(p_{01}-p_{02}+i\epsilon)} 
\over p_{01}-p_{02}+i\epsilon}
A_{\infty}(p_{01},\vec p)B_{\infty}(p_{02},\vec p).
\end{eqnarray}
Expression (\ref{ffff}) is the key for finite-time
thermal field theory.

If $A$ is an operator satisfying assumptions (1) and (2) for
advanced components, we can integrate expression (\ref{ffff})
even further. After closing the $p_{01}$ integration contour in the
lower semiplane, one obtains (if $B$ is an operator satisfying
(1) and (2) for retarded components, one can achieve the same result
by closing the $p_{02}$
integration contour in the upper semiplane):
\begin{eqnarray}\label{fff4}
C_{X_0}(p_0,\vec p)=\int dp_{01}P_{X_0}(p_0,p_{01})
A_{\infty}(p_{01},\vec p)B_{\infty}(p_{01},\vec p).
\end{eqnarray}
This is an extraordinary result: the convolution product of two WTPF's
is a WTPF under conditions (1) and (2).

As expected, in the $X_0=\infty $ limit, Eq.(\ref{fff4}) becomes a
simple product
\begin{eqnarray}\label{fff4i}
\lim_{X_0\rightarrow \infty}C_{X_0}(p_0,\vec p)=
A_{\infty}(p_{0},\vec p)B_{\infty}(p_{0},\vec p).
\end{eqnarray}

At finite $X_0$, Eq.(\ref{fff4}) exhibits
a smearing of energy (as much as it is
necessary to preserve the uncertainty relations).

\subsection{Convolution Product of $n$ Projected Functions }
The product of $n$ two-point functions
is obtained by repeating the above procedure:
\begin{eqnarray}\label{ffffn}
&&C_{X_0}(p_0,\vec p)=\int \prod_{j=1}^{n-1}(dp_{0,j})dp_{0,n}
P_{X_0}(p_0,(p_{0,1}+p_{0,n})/2)
\cr
\nonumber\\
&&
\times\prod_{j=1}^{n-1}\left(A_{j,\infty}(p_{0,j},\vec p)
{1\over 2\pi}{i \over p_{0,j}-p_{0,j+1}+i\epsilon}\right)
e^{-iX_0(p_{0,1}-p_{0,n}+i(n-1)\epsilon)}
A_{n,\infty}(p_{0,n},\vec p).
\end{eqnarray}
For the intermediate products in Eq. (\ref{ffffn}) to hold 
we must require that at least n-1 of the functions
in the product satisfy assumptions (1) and (2). Furthermore, the order of
these functions is important: the retarded functions should
be on the right, the advanced on the left, and
the function that is neither advanced nor retarded in the middle.
However, this is not the order in which
the components appear in the Schwinger-Dyson equation.

If the above requirement is fulfilled, one obtains
(index R for the retarded component, a similar expression
for the advanced component)
\begin{eqnarray}\label{Ij}
&&I_j(p_{0,j-1},p_{0,j+1},\vec p)=\int dp_{0,j}
{1\over 2\pi}{i \over p_{0,j-1}-p_{0,j}+i\epsilon}
A_{R,j,\infty}(p_{0,j},\vec p)
{1\over 2\pi}{i \over p_{0,j}-p_{0,j+1}+i\epsilon}\cr
\nonumber\\&&
=A_{R,j,\infty}(p_{0,j-1},\vec p)
{1\over 2\pi}{i \over p_{0,j-1}-p_{0,j+1}+i\epsilon}.
\end{eqnarray}
Then one finds
\begin{eqnarray}\label{fff4n}
C_{X_0}(p_0,\vec p)=\int dp_{0,1}P_{X_0}(p_0,p_{0,1})
\prod_{j=1}^{n}A_{j,\infty}(p_{0,1},\vec p).
\end{eqnarray}
\section{ Examples of Convolution Products }
\subsection{ Convolution Products of Pole Contributions }
We assume two pole contributions as shown in (\ref{p0}):
$A_{\infty, pole,i}=a_i/(p_0-\bar p_{0,i})$, $i=1,2$.
The product $C=A_1*A_2$ is simple in the cases in which
both contributions are retarded functions, or both are advanced, or
$A_1$ is an advanced and $A_2$ a retarded function.
Then one can simply use (\ref{fff4}) to obtain
\begin{eqnarray}\label{fffra}
C_{X_0}(p_0)=\int dp_{01}P_{X_0}(p_0,p_{01})
{a_1\over p_0-\bar p_{0,1}}{a_2\over p_0-\bar p_{0,2}}.
\end{eqnarray}
The case in which $A_1$ is a retarded and $A_2$ an advanced function (i.e.,
$Im\bar p_{01}<o,~Im\bar p_{02}>0$) requires additional care.
After substituting them into (\ref{ffff}), we choose new variables
$P_0=(p_{01}+p_{02})/2$ and $\Delta_0=p_{01}-p_{02}$, and integrate over
$\Delta_0$ to obtain
\begin{eqnarray}\label{raint}
&&C_{X_0}(p_0)=\int dP_{0}P_{X_0}(p_0,P_{0})
{a_1\over P_0-\bar p_{0,1}}{a_2\over P_0-\bar p_{0,2}}
\cr
\nonumber\\
&&+\int dP_{0}P_{X_0}(p_0,P_{0}){2a_1a_2\over \bar p_{01}+\bar p_{02}-2P_0}
\left({e^{iX_0(2(P_0-\bar p_{01})-i\epsilon)}\over
2(P_0-\bar p_{01})-i\epsilon}
+{e^{-iX_0(2(P_0-\bar p_{02})+i\epsilon)}\over
2(P_0-\bar p_{02})+i\epsilon}\right).
\end{eqnarray}
The first term is formally identical to (\ref{fffra}). The second term
consists of two non-WTPF pieces.

As we shall see in the subsection discussing pinching singularities 
(Sec. 6.4), non-WTPF
contributions appear also in the convolution product of the type 
$G_R*\Omega*G_A$ or, more generally, in the convolution products
containing retarded components 
positioned on the left from the advanced components ($...*A_R*...*B_A*...$).
The non-WTPF terms depend directly on $X_0$, they are
carrying the nontrivial information about the time evolution (i.e., about the
dependence on $X_0$). However, the non-WTPF terms cannot be convoluted further using
Eq. (\ref{ffff}) or (\ref{fff4}).
This fact indicates the natural limits of the applicability
of the methods developed in this paper. 
\subsection{Inverse Propagator and the Equations of Motion}
To define the inverse propagator, we use the results of 
subsection (4.2). We define the restriction of $G_R^{-1}$ on the
subspace of projected functions as 
\begin{eqnarray}\label{gr-1}
G^{-1}_{R,X_0} (p_0,\vec p)=\int dp'_0
P_{X_0}(p_0,p'_0)G^{-1}_{R,\infty}(p'_0,\vec p),
~~~G^{-1}_{R,\infty}(p'_0,\vec p)=i(p^2-m^2+2i\epsilon p_0).
\end{eqnarray}
This integral does not converge in the absolute sense, thus we cannot
calculate the dependence of $G_R^{-1}$ on $X_0$. Nevertheless, we can
apply it from the left to some class of functions.
For example, we can apply it formally to $G_{R,X_0}$: 
$G^{-1}_R*G_R=1$, or written out more explicitly 
\begin{eqnarray}\label{emr}
\int dp'_{0}P_{X_0}(p_0,p'_0)i(p^2-m^2+2ip_0\epsilon)
G_{R,\infty}(p'_0,\vec p)=1.
\end{eqnarray}
This equality is obtained using a 
simple integration over $p_{02}$ in the
expression of the type (\ref{ffff}). We cannot verify
the second identity $G_{R}*G_{R}^{-1}=1$ directly
owing to the divergence
of the integrals, but we can apply it to the projected
function C
\begin{eqnarray}\label{adref}
G_R*G_R^{-1}*C=C,
\end{eqnarray}
under the
only requirement that $C_{\infty}(p_{0},\vec p)$
should satisfy assumptions (1) and (2) in the way a retarded function
does and vanish rapidly enough
at $p_0\rightarrow \infty $ to make the integral over
 $G_{R,\infty}^{-1}(p_{0})C_{\infty}(p_{0})$ convergent.

Equation (\ref{emr}) is the equation of motion for $G_R$. 
In the $X_0\rightarrow\infty$ limit, it reduces to the 
well-known equation for $G_R$.  
For the Keldysh component of the propagator, the equation of motion is given by
\begin{eqnarray}\label{emk}
G^{-1}_R*G_K=G^{-1}_R*(hG_R-hG_A)=0,
\end{eqnarray}
where we have ignored the terms $O(\epsilon)$.
Owing to the presence of the product $G_R^{-1}*hG_A$, this equation cannot be 
verified directly (the integrals diverge). Instead (analogously to the case of
the product $G_R*G_R^{-1}$), one multiplies it from the left by the function C,
which vanishes rapidly at $p_o\rightarrow \infty $ and satisfies assumptions 
(1) and (2) in the way 
an advanced function does.
\subsection{Resummed Schwinger-Dyson Series }
We write the Schwinger-Dyson equations in the form
\begin{eqnarray}\label{sde}
&&{\cal G}_R=G_R+iG_R*\Sigma_R*{\cal G}_R,~~
{\cal G}_A=G_A+iG_A*\Sigma_A*{\cal G}_A,
\cr
\nonumber\\
&&{\cal G}_K=G_K+iG_R*\Sigma_K*{\cal G}_A
+iG_K*\Sigma_A*{\cal G}_A+iG_R*\Sigma_R*{\cal G}_K.
\end{eqnarray}
The formal solution (where all products are convolution products and
the operators are kept in the proper order) is
\begin{eqnarray}\label{sdsol1}
{\cal G}_R=G_R*(1-i\Sigma_R*G_R)^{-1},~~
{\cal G}_A=G_A*(1-i\Sigma_A*G_A)^{-1},
\end{eqnarray}
\begin{eqnarray}\label{sdsol2}
{\cal G}_K={\cal G}_R
*(h(p_0,\omega_p)(G^{-1}_A-G^{-1}_R)+i\Omega)*{\cal G}_A.
\end{eqnarray}

To use the formal solution of the Schwinger-Dyson series, we assume that the
functions $G_{R(A)}$, ${\cal G}_{R(A)}$, and $\Sigma_{R(A)}$
satisfy requirements (1) and (2) for the
retarded components in the upper and for the advanced in the lower
semiplane.

This assumption deserves a few comments:
For the retarded (advanced) bare propagators, our assumption is valid.

If the retarded component is real between the cuts on the
part of the real axis,
the Schwartz theorem tells us that assumptions (1) and (2)
valid in the upper semiplane are also valid
in the lower semiplane of the first Riemann sheet.

At equilibrium, perturbation theory yields the full propagator as
a set of Fourier coefficients. The analytic continuation in the energy
plane is not unique. This freedom is used to choose an analytic
continuation that satisfies requirements (1) and (2) defined
in Sec. 3. The positivity property of the spectral density then implies
that the propagator
has neither zeroes nor poles off the real axis$^{\cite{landsman}}$.
A further implication is that the exact self-energy
$\Sigma_{R}(p_0,\vec p)$ at equilibrium also satisfies
the properties (1) and (2). This is not guaranteed
for approximate expressions for self-energy.

In the formal solution of the retarded
propagator, the factors $G_R$ and $\Sigma_R$ alternate regularly. This
fact can improve the convergence properties of some integrals.

Now it is easy to write down the resummed Schwinger-Dyson series for
the retarded (advanced) propagator (with any exact self-energy
obtained by the perturbation expansion that satisfies our assumptions).
In terms of the corresponding
propagator calculated at $X_0=\infty$:
\begin{eqnarray}\label{fff5}
{\cal G}_{R(A),X_0}(p_0,\vec p)=\int dp_{01}P_{X_0}(p_0,p_{01})
{\cal G}_{R(A),\infty}(p_{01},\vec p),
\end{eqnarray}
where
\begin{eqnarray}\label{rafff5}
{\cal G}_{R(A),\infty}(p_{01},\vec p)={G_{R(A),\infty}(p_{01},\vec p)\over
1-i\Sigma_{R(A),\infty}(p_{01},\vec p)G_{R(A),\infty}(p_{01},\vec p)}.
\end{eqnarray}
Starting from Eq. (\ref{rafff5}) one obtains 
the HTL-resummed$^{\cite{carrington,defu}}$ retarded
(advanced) component of the propagator 
without use of the gradient expansion.

Some more work is necessary to calculate the Keldysh component.
Now,in addition to individual terms, the sum
${\cal G}_{R,\infty}(p_0,\vec p)$ should also satisfy (1) and (2)
i.e., the imaginary part of $\Sigma_{R,\infty}$ should be negative.

However, there is a possibility that Im$\Sigma_R$ is positive in some
kinematical region. Then the resummed Schwinger-Dyson equation
for a retarded component can create the pole in the upper semiplane.
However, this case is very questionable: one sums infinitely many
retarded functions (i.e., the functions which vanish at
$t<t'$) and obtains the function which is not retarded (i.e.,
nonzero at $t<t'$). Such cases are usually classified as
pathology$^{\cite{henning2,henning}}$).
At this point one should cautiously consider the use of the
"physical" gauge$^{\cite{rl}}$, in order to prevent
eventual gauge artifacts.

Some indication that, in some cases, ${\cal G}_{R,\infty}(p_0,\vec p)$
does satisfy assumptions (1) and (2) comes from the HTL limit.
Indeed, at equilibrium, the HTL limit of
${\cal G}_{R,\infty}(p_{01},\vec p)$
must satisfy (1) and (2),
as it is easy to verify. As the properties of density functions
enter only through the thermal mass and the position of isolated poles,
the same must be true of any distribution allowing the HTL
approximation.

Owing to the fact that the Keldysh component of self-energy does
not satisfy the analyticity assumptions (1) and (2),
we can only try to integrate expression (\ref{sdsol2}) 
using approximate and numerical methods.

However, it is possible that $\Sigma_K $ can be
decomposed into two pieces satisfying assumptions (1) and (2)
as retarded and as advanced functions, respectively:
$\Sigma_K= -\Sigma_{K,R}+\Sigma_{K,A}$. For example, this happens in the case
of one-loop self-energy.
Then Eq. (\ref{sdsol2}) becomes
\begin{eqnarray}\label{sdsol2s}
{\cal G}_K={\cal G}_R*(h(p_0,\omega_p)G^{-1}_A+i\Sigma_{K,A})*{\cal G}_A
-i{\cal G}_R*(h(p_0,\omega_p)G^{-1}_R+i\Sigma_{K,R})*{\cal G}_A.
\end{eqnarray}
Owing to the fact that the functions ${\cal G}_R$ and ${\cal G}_A$ are not
singular in the point $p_0=\pm\omega_p$,
the terms containing $G_R^{-1}$ 
and $G_A^{-1}$ cancel mutually. 
As one of the remaining convolutions includes factors of the same type
(RR or AA), we are left with a single convolution multiplication. 
This convolution contains
neither the advanced first factor nor the retarded second factor; thus, 
in general, it cannot be 
worked out in a simple way, and it will contain non-WTPF 
contributions. However, it may be 
performed at least numerically. 

The appearance of the non-WTPF contributions signals 
stepping out of the space of
projected functions. Indeed, the calculation
of the more complex diagrams, containing subdiagrams resummed into
${\cal G}_K$, will not enjoy advantages of the presented calculus.

Finally, we note here that the calculation starting with Eqs. (\ref{sde})
and ending with (\ref{fff5}) - (\ref{sdsol2s}) cannot be performed with the
true (i.e., calculated, in some miraculous way, to all orders) $\Sigma_K$,
$\Sigma_R$, and $\Sigma_A$. Indeed, we have anticipated that the true
$\Sigma_K$, $\Sigma_R$, and $\Sigma_A$ contain non-WTPF terms, and thus one cannot use
instead of $G_R^{-1}$ and $G_A^{-1}$ their restrictions to the subspace
of projected functions. Using the gradient expansion (under large-$X_0$
assumption) one obtains familiar equations of motion for the Green functions
of interacting fields$^{\cite{keldysh,kadanoff,danielewicz,bi}}$. However, 
the advantage of the presented
calculus will be observed in the properties of collision integrals, where
one can expect considerable simplifications and the possibility of
evaluating contributions of more complex diagrams. A more complete discussion of
collision integrals is out of scope of the present paper, and we hope to publish it elsewhere.
\subsection{Pinching singularities}
The pinchlike contribution to the Keldysh component of the resummed
propagator is expressed as$^{\cite{id}}$ (we treat only the scalar case)
\begin{eqnarray}\label{sdsol2p}
G_{Kp}=iG_R*\left(-\bar\Sigma_{K,R}+\bar\Sigma_{K,A}\right)*G_A,
\end{eqnarray}
where we have introduced the short notation
$\bar \Sigma_{K,R(A)}=h(p_0,\omega_p)\Sigma_{R(A)}+\Sigma_{K,R(A)}$.

Similarly as in the case of resummed contributions, we can
perform convolution between alike components (RR or AA).
Then one can integrate the terms containing $\Sigma_R$ and $\Sigma_{K,R}$
with respect to $p_{02}$, and the terms containing $\Sigma_A$
and $\Sigma_{K,A}$ with respect to $p_{01}$.
The result is intriguing:
\begin{eqnarray}\label{sdsol2pf}
&&G_{Kp,X_0}(p_0,\vec p)=-\int dp_{01}P_{X_0}(p_0,p_{01})
{1\over p_{01}^2-\omega_{p}^2+2i\epsilon p_{01}}i\bar \Sigma_K
{1\over p_{01}^2-\omega_{p}^2-2i\epsilon p_{01}}
\cr
\nonumber\\
&&-{1\over 2\omega_p}\sum_{\lambda=-1}^1\lambda\left(
\int dp_{01}{e^{2iX_0(p_0-p_{01})}-e^{-2iX_0(p_0-\lambda\omega_p)}
\over i\pi(2p_0-p_{01}-\lambda\omega_p)}
\bar\Sigma_{K,R}(p_{01}+i\epsilon,\vec p)
{1\over p_{01}^2-\omega_{p}^2+2i\epsilon p_{01}}\right.
\cr
\nonumber\\
&&+\left.
\int dp_{01}{e^{-2iX_0(p_0-p_{01})}-e^{2iX_0(p_0-\lambda\omega_p)}
\over i\pi(2p_0-p_{01}-\lambda\omega_p)}
\bar\Sigma_{K,A}(p_{01}-i\epsilon,\vec p)
{1\over p_{01}^2-\omega_{p}^2-2i\epsilon p_{01}}
\right).
\end{eqnarray}
The first term in Eq. (\ref{sdsol2pf}) is a projected function (WTPF)
that becomes the usual pinchlike term in the $X_0\rightarrow \infty $ limit.
It is this contribution to which the conclusions$^{\cite{id}}$
about the cancellation of pinching singularities apply .
However, the other terms are of non-WTPF nature; contrary to the case of
the product of simple pole terms, the discontinuity along the real axis
appearing in the functions $\Sigma_{R(A)}$ and $\Sigma_{K,R(A)}$ now prevents
the vanishing of these terms.

A full discussion on pinching singularities in the
finite-time-after-switching formulation requires more efforts and
we hope to publish it elsewhere.

\section{ Modifications of the Feynman rules}
In this section, in the framework of the generic field theory with bosons
and fermions, we discuss the changes of Feynman rules that are
due to the ``finite time'' 
assumption. We further analyze the diagrams  
with respect to the question of energy 
nonconservation. Indeed, we find that this feature appears together with the
 non-WTPF contributions.  

The calculations performed so far already contain all of the
modifications of the Feynman rules required by the finite-$t_i$ assumption.
In coordinate space, the only modification is that the bare propagators
[Eqs. (\ref{nonigrr}) and (\ref{nonigrk})]
are limited by $0<x_0$ and $0<y_0$; thus they
are projected functions.
In energy-momentum space, the above change reflects in the change of
propagators, vertices, and the overall factor.

To transform to energy-momentum space, we choose some vertex $j$,
arrange the orientation so that all lines $i$ become outgoing, and
use the propagators represented by Eqs. (\ref{nonigrr}),
(\ref{nonigrk}), and (\ref{fnonigrkr}) (the $p_i$ momentum is joined
to the line $i$). Exponentials
attached to $x_j$ are easily integrated with the help of
Eq.(\ref{zint}):
\begin{eqnarray}\label{xjint}
{1\over 2\pi}\int_0^{\infty}dx_je^{-ix_j(\sum_ip_i-i\epsilon})=
{i\over 2\pi(-\sum_ip_i+i\epsilon)}.
\end{eqnarray}

After performing this integration, instead of the bare propagators
we obtain their
$X_0\rightarrow \infty $ limits [Eq. (\ref{fnonigrkr})], which are
the familiar propagators of the usual ($t_i\rightarrow -\infty $)
theory.

At the vertices the usual energy-conserving $\delta(\sum_i p_{0i})$
is substituted by
$i(2\pi)^{-1}(-\sum_ip_{0i}+i\epsilon)^{-1}$.

Under the momentum integrals there is a
leftover factor at the vertices $j_A$ (by subscript $A$ we indicate
that $j_A$ are vertices with amputated legs):
\begin{eqnarray}\label{xjaint}
e^{-i\sum_{j_A}x_{j_A}(\sum_{i_{j_A}}\lambda_{i_{j_A}}p_{i_{j_A}})},
\end{eqnarray}
where $\lambda=\pm$ depends on whether the corresponding momentum is outgoing of or
incoming to the vertex $j_A$, and $i_{j_A}$ is running through the
nonamputated lines.

The overall factor in the case of two-point functions
is treated in a simple way: introduce a slow Wigner variable as the 
average over the times of boundary vertices,
and the relative time [Eq. (\ref{wv})].
Finally, one can Fourier transform over the relative time.
There emerges an overall energy-smearing factor
$P_{X_0}(p_0,p'_{0})$ for two-point functions and similarly
for n-point functions. In the case of n-point functions, the choice of
variables is large and might not be unique; namely, depending on the
diagram calculated, one chooses the most appropriate set of variables.

The overall factor takes care of
uncertainty relations: the larger the elapsed ``time'' $X_0 $, the
smaller the energy smearing.

In the vertex factor the energy is
not explicitly conserved. This energy nonconservation is, through the
uncertainty relations, related to the finiteness of $X_0$. In the limit of 
infinite $X_0$,
energy conservation is recovered. Here we want to argue that for some choice
of propagators entering the vertex, the energy is conserved explicitly.
To see this conservation,
assume, for a moment, that at least one of the unspecified 
propagators ($D$, $G$, and $S$)
related to the chosen vertex, say $D_\infty$,
is a retarded function, $D_R$. In this case,
one can integrate over $q_0$, close the integration path from above
(owing to
$e^{iX_0(p_{o,j}+i\epsilon)}$, closing from below is out of question), 
and collect
the contributions from singularities. If there are no singularities
(and we know that conditions (1) and (2) are valid for bare
propagators), one just obtains the
energy-conservation condition $\delta(\sum_ip_{0i})$.
The same is achieved with the outgoing momenta and advanced components
of the propagator with closing the integration path from below.
Now we are going to show that, indeed, 
one of these possibilities is
realized at each vertex.

Each individual denominator
$(\sum_ip_{0i}-i\epsilon)^{-1}$ (the lines are all oriented out)
can be easily
integrated.
To demonstrate this, we have to sum over the indices of the
corresponding vertex. We rename the basis
$(i,j), i,j=1,2$ into $[\mu,\nu]$, where $\mu,\nu=-1$ correspond to
$i,j=2$, and $\mu,\nu=1$ correspond to $i,j=1$.
Then we find the relations of the type (we assume a three-field
vertex, but the proof extends
easily to any number of fields):
\begin{eqnarray}\label{munu}
D_{[\mu,\nu]}={1\over 2}(D_K-\mu D_R-\nu D_A).
\end{eqnarray}
The sum over the indices in the chosen vertex (S -
, D -, G - propagators of the outgoing lines;
the factor $\mu$ for the negative coupling of the 
vertex to which the index-2 ends of the propagators are attached) is
\begin{eqnarray}\label{2munu}
&&\sum_\mu \mu S_{[\mu,\lambda]}D_{[\mu,\rho]}G_{[\mu,\nu]}=
{1\over 4}\biggl(S_RD_RG_R+(S_K+\lambda S_A)(D_K+\rho D_A)G_R
\cr
\nonumber\\
&&+(S_K+\lambda S_A)D_R(G_K+\nu G_A)
+S_R(D_K+\rho D_A)(G_K+\nu G_A)\biggr).
\end{eqnarray}
Expression (\ref{2munu})
contains only terms  including at least one retarded propagator:
$S_R$, or $D_R$, or $G_R$.

Thus one
can integrate the terms separately and find that the factor
$(\sum_ip_{0i}-i\epsilon)^{-1}$ is effectively replaced by
$i\pi\delta(\sum_ip_{0i})$.

As there is nothing special at this vertex
(the indices $\lambda, \rho, \nu$ remain unspecified),
one may conclude that this is a general feature. Nevertheless,
one should do it very cautiously, step by step, while
problems may appear at some degree of complexity.
Then, as seen in Eqs. (\ref{ffff}) and (\ref{ffffn}),
we find a new element in addition to the energy denominator
 $(-p_{0j}+p_{0j+1}-i\epsilon)^{-1}$. One obtains the extra factor
 $e^{-iX_0(p_{0j}-p_{0j+1}-i\epsilon)}$.
With the help of this factor, even the contributions from the poles
of the retarded component in the upper semiplane will decay
exponentially with the time $X_0$.

However, the diagrams with resummed self-energy subdiagrams are
particularly sensitive. In this case, one is strongly advised to
undertake an intermediate step: to Fourier transform the two-point
function with respect to the relative time, to investigate the analytic
structure, and then to perform the multiplication of two-point functions.
Owing to the cuts in ${\cal G}_{R(A)}$  and to the non-WTPF contributions
to ${\cal G}_K$, it is likely that
this is just the point after which we have to live without
the advantages of projected functions.
\section{Summary}
We consider out of equilibrium thermal field theories with switching on
the interaction occurring at finite time ($t_i=0$).

We study Wigner transforms (also in the relative time $s_0$) of
two-point functions.
To develop a calculation scheme based on first principles,
we define a very useful concept of projected functions: a two-point
function with the property that it is zero for $x_o<t_i$ and for
$y_0<t_i$; for $t_i<x_0$ and $t_i<y_0$, the function depends only on
$x_0-y_0$. We find that many important functions are of this type:
bare propagators, one-loop self-energies, resummed Schwinger-Dyson
series with one-loop self-energies for the case of
retarded and advanced components of the propagator, etc.
The properties of the Wigner transforms
in the $X_0\rightarrow \infty $ limit are particularly simple
if they satisfy these analyticity assumptions: (1) The function of $p_0$ is
analytic above the real axis (for a retarded component, but
below it for an advanced component).
(2) The function goes to zero
as $|p_0|$ approaches infinity in the
upper (lower) semiplane. We find that these assumptions are very natural
at low orders of the perturbation expansion.
The convolution product of projected functions is remarkably simple,
much simpler than what one would expect from the gradient
expansion.

The Schwinger-Dyson series, with bare propagators and self-energies
being projected functions satisfying assumptions (1) and (2), is
resummed in closed form without the need for the gradient expansion.
The calculation of the resummed Keldysh component is simplified to
a double (and under a certain analyticity assumption, to a single)
convolution product. This contribution signals the stepping out of the
comfortable space of projected functions.

The Feynman diagram technique is reformulated: there is no explicit
energy conservation at vertices, there is an overall energy-smearing
factor taking care of the finite elapsed time ($X_0$) and the
uncertainty relations.

The relation between the amplitudes (valid at low orders of the
perturbation expansion) of the theory with switching on the
interaction in the remote past
and the theory with finite switching-on time, enables one to rederive
the results such as cancellation of pinching singularities,
cancellation of collinear and infrared singularities, HTL resummation,
etc. Previously, these
results were considered applicable only to lowest-order
contributions in the gradient expansion.

The question arises whether higher-order contributions also
remain within the space of projected functions satisfying assumptions
(1) and (2). The answer depends on the eventual positivity of Im$\Sigma_R$,
explicitly time-dependent perturbation, and the appearance of the
one-loop approximated or resummed Keldysh component.
The positive Im$\Sigma_R$ can create the pole in the upper semiplane
in the resummed Schwinger-Dyson series.
However, this case is very questionable: one sums infinitely many
retarded functions (i.e., functions which vanish for
$t<t'$) and obtains the function which is not retarded (i.e.,
nonzero at $t<t'$). Such cases are usually classified as
pathology$^{\cite{henning,henning2}}$).
The way of breaking the scheme explicitly
is to introduce direct time dependence through
time-dependent perturbation, or through the background field which
depends directly on time: $\psi(x,t)=\phi(t)+\bar \psi(x,t)$.
In this way, one obtains the two-point functions which are not
projected functions.
A natural step out of
the space of projected functions occurs
in the calculation of the resummed Keldysh component of the propagator.
The appearance of the non-WTPT contributions signals that the calculation
of the more complex diagrams containing, subdiagrams resummed into
${\cal G}_K$, will not enjoy advantages of the presented calculus.

{\large \bf Acknowledgments}
 I am especially indebted to Rolf Baier for
turning my attention to the obstacle of gradient expansion and
for useful remarks, and to Jean Cleymans for many helpful discussions.
I also acknowledge
financial support from the EU under contract
No. CI1$^*$-CT91-0893(HSMU) and from the A. von Humboldt Fondation.
This work was supported by the Ministry of Science and Technology
of the Republic of Croatia under contract No. 00980102.

\newpage
{\Large \bf Figure Captions}

Fig. 1: Finite switching-on time integration path.

\end{document}